\documentclass[amsmath,amssymb,aps,showpacs,floatfix,prl,twocolumn]{revtex4-1}
\usepackage{graphicx}
\usepackage{bm}
\usepackage{hyperref}
\usepackage{color}
\begin{document}
\newcommand\psone{p_1(n_1,n_2;t)} 
\newcommand\pstwo{p_2(n_1,n_2;t)} 
\title{Non-Markovian Models of Blocking in Concurrent and Countercurrent Flows}
\author{A. Gabrielli$^1$, J. Talbot$^2$ and P. Viot$^2$}
\affiliation{
$^1$Instituto dei Sistemi Complessi (ISC) - CNR, UOS "Sapienza", Physics Department, University "Sapienza" of
Rome, Piazzale Aldo Moro 2, 00185 - Rome, Italy\\
$^2$Laboratoire de Physique Th\'eorique de la Mati\`ere Condens\'ee, UPMC, CNRS  UMR 7600,
 4, place Jussieu, 75252 Paris Cedex 05, France
}

\date{\today}
\begin{abstract}
We investigate models in which blocking can interrupt a particulate flow process at any time. 
Filtration, and flow in micro/nano-channels and traffic flow are examples of such processes. 
We first consider concurrent flow models where particles enter a channel randomly. 
If at any time two particles are simultaneously present in the channel, failure occurs. 
The key quantities are the survival probability
and the distribution of the number of particles that pass before failure.  
We then consider a counterflow model with two opposing Poisson streams. There is no
restriction on the number of particles passing in the same direction, but blockage occurs if, at any time, 
two opposing particles are simultaneously present in the passage. 
\end{abstract}

\maketitle
{\it Introduction.}
Processes involving the flow of particles through channels may entail blocking or failure. 
A good example for a {\it concurrent} flow is provided by the industrially important process 
of filtration\cite{Hampton19931601,Schwartz1993,Lee1996,Redner2000,PhysRevLett.98.114502}. 
In particular, the model of Roussel et al.  
\cite{PhysRevLett.98.114502} successfully accounted for experimental data by assuming that
clogging may occur when two grains are simultaneously present in the vicinity of a mesh hole, even though isolated grains are small 
enough to pass through the holes. A conceptually similar situation is a flimsy bridge that can only support 
the weight of one car at a time. If ever two cars are on the bridge at the same time, it collapses. 

A second class of processes involves two {\it counterflowing} streams of particles. For example, in remote areas many of the roads are 
single-track. Two approaching vehicles cannot pass each other except at rather infrequent, and short, passing places. In this 
situation we would like to know the failure probability of finding two opposing cars in the stretch of road between two passing places.

Many traffic models based on lattice gases have been proposed
\cite{Helbing1999,Helbing2001,Schiffmann2010,Moussa2012,Hilhorst2012a,Appert-Rolland2010,Appert-Rolland2011,Ezaki2012}
including the totally
asymmetric simple exclusion processes (TASEP) \cite{derrida1992,Evans1995} and related models \cite{Reuveni2012}.
The so-called bridge models\cite{Popkov2008,Evans1995,Evans1995a,Godreche1995,Grokinsky2007,Willmann2005,jelic2012} consider two TASEP 
processes with oppositely directed flows, but allow exchange of particles on the bridge. 

Similar processes are also found in numerous biological applications involving channels. Examples include 
bidirectional macromolecular flow in microchannels \cite{Champagne2010}, ion channels 
that can be clogged by toxins or medicines \cite{kapon2008,Kim2008,Twiner2012}, and  the antibiotic gramicidin that forms univalent cation-selective
channels of  0.4nm diameter in phospholipid bilayer membranes. The transport of ions and water throughout most of the channel length is by a single file
process; that is, cations and water molecules cannot pass each other within the channel \cite{Finkelstein1981}.

In this Letter we propose, and obtain exact solutions for, stochastic models in which particle flow in a channel can be instantaneously interrupted
by a clogging  event. The quantities of interest are the probability of blockage (failure) as a function of time and the 
final outcome, i.e. the number and type
of particles that get through the channel before blockage occurs.
These models are complementary to the lattice gas models in that they are continuous in both space and time and are most appropriate for low density flows.


\begin{figure}[ht]
\begin{center}
\includegraphics[width=8cm]{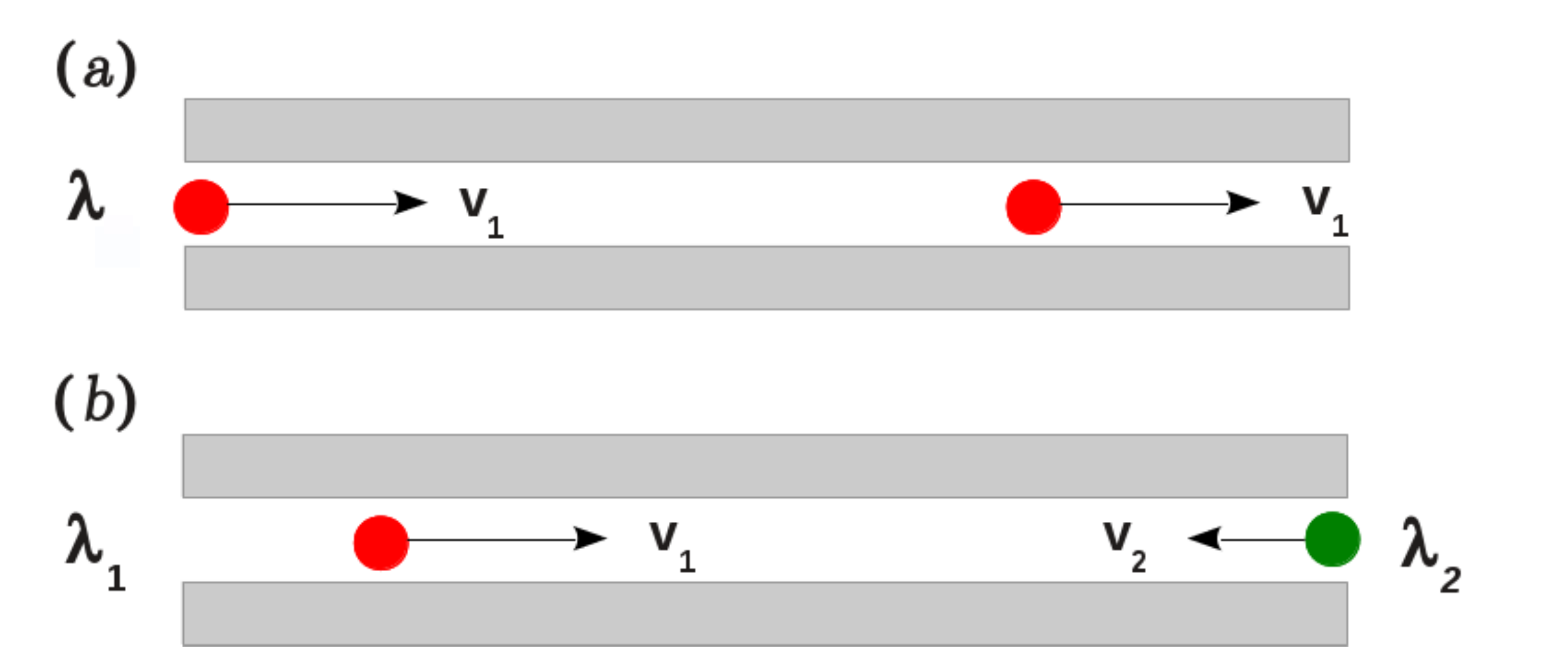}
\end{center}
\caption{(a) Concurrent flow model: Particles enter 
the left hand side of a channel of length
$L$ at a (mean) rate $\lambda$. 
Blockage occurs when two particles are simultaneously present in the channel.
(b) Counterflow model: Two opposing streams of particles
enter the left and right hand sides of the channel at
rates $\lambda_1$ and $\lambda_2$, respectively. 
Blockage occurs whenever two opposing particles are simultaneously present in the channel 
but there is no constraint on the number of particles moving in the same direction.}\label{fig:channel}
\end{figure}

{\it Concurrent flow model.} Particles enter a passage of length $L$ according to a homogeneous Poisson process where
\begin{equation}\label{eq:poisson}
 P_n(t)=\frac{(\lambda t)^n}{n!}\exp(-\lambda t)
\end{equation}
gives the probability that $n$ particles enter the passage in the time interval $(0,t)$. 
We assume that all particles move with constant velocity $v$ so that the transit time, $\tau=L/v$, is constant.
Blockage (failure) occurs at the instant when two particles are present in the channel at the same time  
(see Fig.~\ref{fig:channel} a). This leads us to consider the survival probability  
$p_s(t)$,  the probability that blockage (failure) does not occur in the time interval $(0,t)$. Clearly, $p_s(0)=1$ and $p_s(\infty)=0$. 
The probability that blocking 
occurs between time $t$ and $t+dt$ is given by $f(t)dt$ where $f(t) = -dp_s(t)/dt$.

To solve the model we introduce the $n$ particle survival probability $q_s(n,t)$ 
which denotes the joint probability of surviving up to $t$ and that $n$ particles have entered the passage during this time. 
The survival probability is simply $p_s(t)=\sum_{n\ge 0} q_s(n,t)$. 
The evolution of the $q_s(n,t)$ is given by:
\begin{equation}
\begin{cases}
& \frac{d q_s(0,t)}{dt}=-\lambda q_s(0,t)\\
& \frac{d q_s(1,t)}{dt}=\lambda q_s(0,t)-\lambda q_s(1,t) \\ 
& \frac{d q_s(n,t)}{dt}=\lambda q_s(n-1,t-\tau)e^{-\lambda \tau}-\lambda q_s(n,t) \mbox{\vspace{2cm},\; $n\geq 2$}  
\end{cases}\label{eq:cars}
\end{equation}
The final equation implements a non-Markovian constraint: in passing from the state \{"not blocked", $n-1$\} to the state \{"not blocked", $n$\} it is necessary that 
no particle enter in the previous time interval $(t-\tau, t)$ and that a single particle enter in the time interval $t$ to $t+dt$. These probabilities are given by $e^{-\lambda\tau}$ and $\lambda dt$, 
respectively. By introducing the generating function we obtain, as detailed in the Supplementary Material (SM),
\begin{equation}\label{eq:ps5}
p_s(t)=\left(1+\sum_{n=0}^\infty\theta(t-n\tau)\frac{(\lambda(t-n\tau))^{n+1}}{(n+1)!}\right) e^{-\lambda t}
\end{equation}
where $\theta(t)$ is the Heaviside function.

\begin{figure}[th]
\begin{center}
\includegraphics[width=7cm]{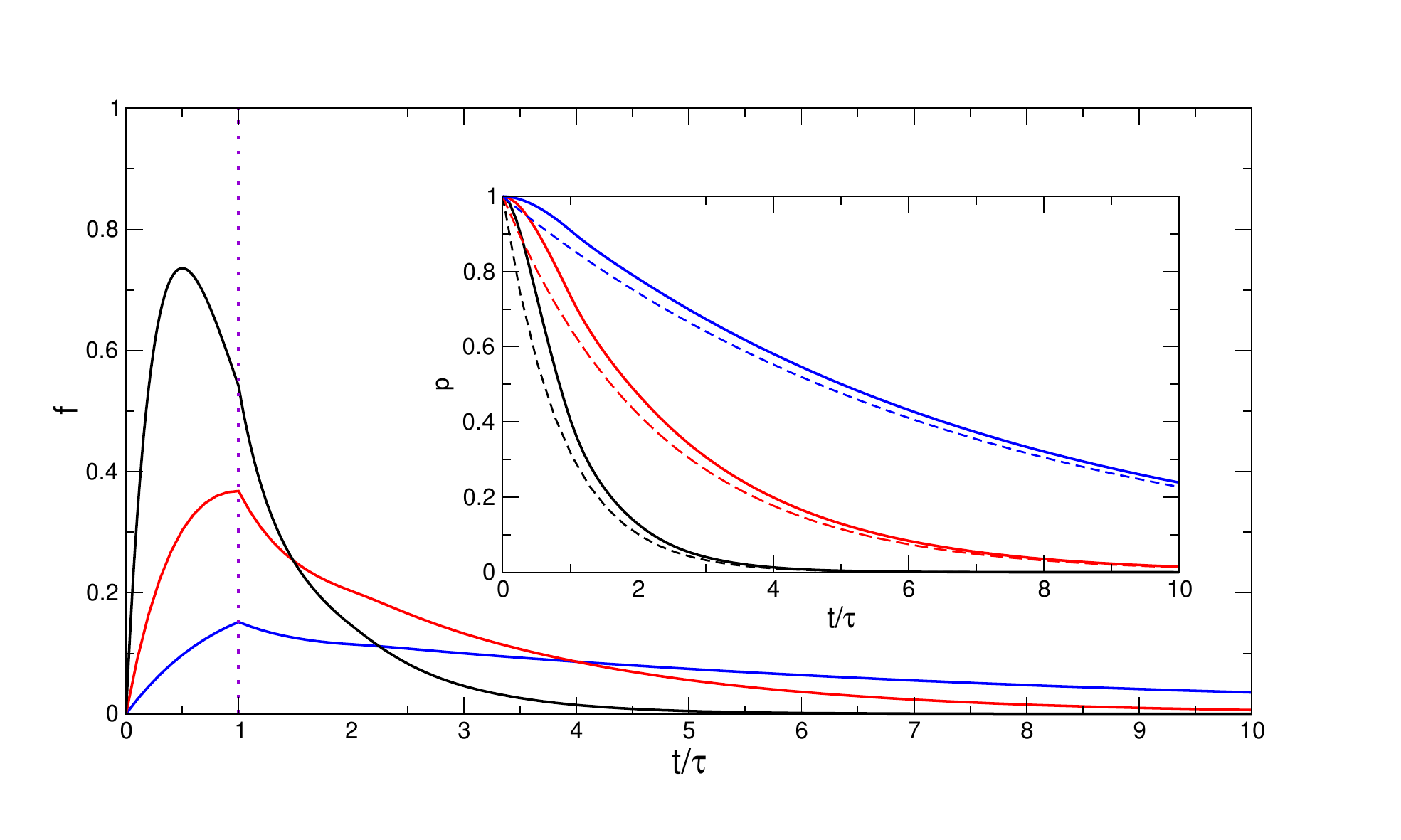}
\end{center}
\caption{Probability that blockage occurs as a function of time for $\lambda\tau=2,1,0.5$ (most peaked to least peaked). 
Cusps are present at $t/\tau=1$ (vertical dotted line).
The inset shows the survival probability for the same parameters together with the asymptotic approximation, 
Eq. (\ref{eq:asym1}) (dashed lines).}\label{fig:fps}
\end{figure}

The long time behavior of the survival probability can be obtained by approximating the sum in 
Eq. (\ref{eq:ps5}) as an integral 
and evaluating it using the Saddle-Point (Laplace) method. The result is
\begin{equation}\label{eq:asym1}
  p_s(t)\sim e^{-\left(\lambda -\frac{L_W(\lambda \tau)}{\tau}\right)t}
\end{equation}
where $L_W(x)$ is the Lambert-W function (see Sect.~1.2 of the SM). 
From its small $x$ behavior one can deduce that, when $\lambda\tau\ll 1$, the  exponent of the exponential decay, 
$\lambda-\frac{L_W(\lambda \tau)}{\tau}\simeq \lambda^2 \tau$,  
depends nonlinearly on the rate $\lambda$. This complexity arises from the large possible number of event sequences before failure. 

The mean survival time is given by
\begin{equation}
 \langle t\rangle = \int_0^{\infty}p_s(t)dt = \frac{2e^{\lambda\tau}-1}{\lambda(e^{\lambda\tau}-1)}
\end{equation}
and is consistent with Eq.~(\ref{eq:asym1}) when $\lambda\tau\ll 1$.

Figure \ref{fig:fps} illustrates the time dependent properties of the concurrent flow model. 
The curves showing the probability of failure at time $t$, $f(t)$, exhibit a cusp 
at $t=\tau$. This is the first time at which particles that have entered previously
can exit the channel (which is empty at $t=0$), leading to a rapid decrease in the probability of blockage. The intensity of the cusp depends on $\lambda$, 
 ($df/dt|_{\tau^-}-df/dt|_{\tau^+} = \lambda^2e^{-\lambda \tau}$), and is less pronounced for large $\lambda$ as a second particle is more likely to enter
soon after the first, causing blockage. The inset shows the survival probability and confirms the accuracy of the asymptotic 
expression, Eq. (\ref{eq:asym1}). 

A further quantity of interest is the
distribution of number of particles that exit the channel before blockage occurs. If $n$ particles have entered the passage
at failure, the number that have successfully traversed is $m=n-2$. At least two particles must enter before failure can occur.
Let $h(m)$ denote the probability that when failure occurs $m$ particles have exited. If $\Delta t_i$ denotes the time interval 
between the entry of the $i^{th}$ and the $(i+1)^{th}$ particle, then
\begin{equation}
 h(m)=\left[\prod_{i=1}^m {\rm Pr}(\Delta t_i>\tau)\right]{\rm Pr}(\Delta t_{m+1}<\tau)
\end{equation}
Using that ${\rm Pr}(\Delta t>\tau)=e^{-\lambda\tau}$ and ${\rm Pr}(\Delta t<\tau)=1-e^{-\lambda\tau}$ we find
\begin{equation}
 h(m)=e^{-m\lambda\tau}(1-e^{-\lambda\tau})
\end{equation}

\begin{figure}[t]
\begin{center}
\includegraphics[width=8cm]{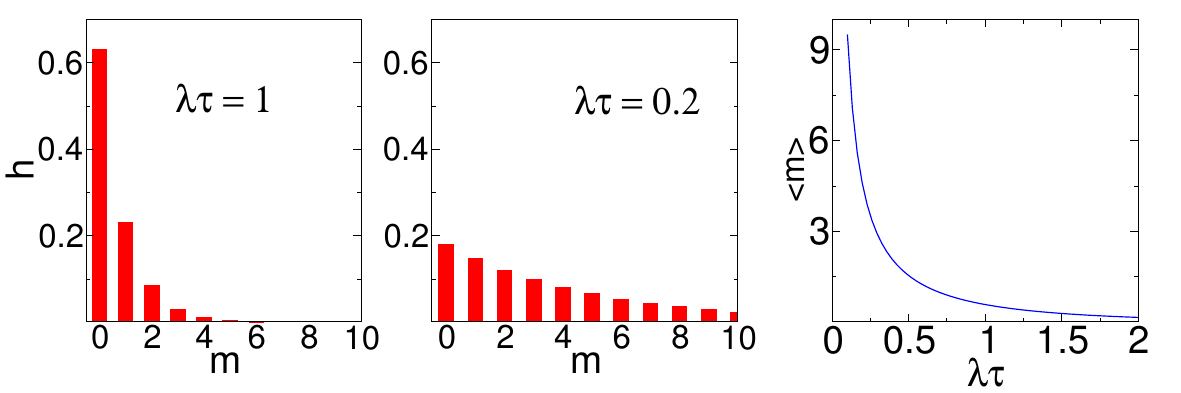}
\end{center}
\caption{Distribution of the number of particles that pass before blockage occurs for different values of $\lambda\tau$ (left and middle figures). 
The right graph shows the average number
as a function of the same parameter.}\label{fig:hdist}
\end{figure}
The most probable situation is that no particles pass before
failure for all values of $\lambda$. 
The mean number that pass before failure is
\begin{equation}\label{eq:nbar}
 \left\langle m\right\rangle = \frac{1}{e^{\lambda\tau}-1}
\end{equation}
which has the expected asymptotic behavior: $\left\langle m\right\rangle \rightarrow e^{-\lambda\tau}$ for $\lambda\tau$ large and 
$\left\langle m\right\rangle \rightarrow (\lambda\tau)^{-1}$ for $\lambda\tau$ small.
Figure \ref{fig:hdist} illustrates that with decreasing $\lambda\tau$ the difference between the mean, $\left<m\right>$, of $m$ and its most probable value (always 0) 
increases and $h(m)$ becomes flatter.

The above results can be generalized for a distribution of transit times. If $\psi(\tau)$ is the normalized distribution of
transit times and assuming that $\lambda$ is constant, the mean survival time is
\begin{equation}
 \langle t\rangle = \frac{2-\tilde{\psi}(\lambda)}{\lambda(1-\tilde{\psi}(\lambda))}
\end{equation}
and the mean number of particles that pass before failure is
\begin{equation}
  \langle m\rangle = \frac{\tilde{\psi}(\lambda)}{1-\tilde{\psi}(\lambda)}
\end{equation}
where tilde denotes the Laplace transform, $\tilde{\psi}(\lambda)=\int_0^{\infty}e^{-\lambda\tau}\psi(\tau)d\tau$.

{\it Counterflow model.} In this model particles of type 1 enter the channel of length $L$ at the left at a rate $\lambda_1$ and move
towards the right at a speed $v_1$.
Particles of type $2$ enter at the right at a rate $\lambda_2$ and move to the left at speed $v_2$:  
(see Fig.~\ref{fig:channel}b)
The transit times are $\tau_1=L/v_1$ and $\tau_2=L/v_2$,
respectively.  We assume that particles enter according to a Poissonian distribution so that
the probability that $n_1$ ($n_2$) particles of type 1 (2) enter the left (right) side in the interval
$(0,t)$ is given by:
\begin{equation}\label{eq:poisson2}
 P_{n_i}(t)=\frac{(\lambda_i t)^{n_i}}{n_i!}e^{-\lambda_i t}\;\;\;{\rm with}\;i=1,2
\end{equation}
A blockage occurs if, at any time, particles of both species are present in the
channel. Before this situation arises an arbitrary number of particles can transit the passage in both directions.
If $\lambda_i\tau_i < 1$ the average time interval between entry of particles of type $i$ is longer than the transit time. 
If, on the other hand,  $\lambda_i\tau_i > 1$, a backlog of particles of type $i$ is likely to be present. Thus the former situation is more relevant physically. 

We now outline the solution method.  The device that allows us to obtain an analytical solution in this case is the introduction of functions
$p_k(n_1,n_2;t)$ that denote the probability that the system has survived until time $t$ and $n_1$  particles of type 1 and $n_2$ of type 2 have entered the passage
{\em and the last particle to enter the passage was of type $k=1,2$}. This choice provides a complete partition of the event space into disjoint events 
allowing us to write $p(n_1,n_2;t)=p_1(n_1,n_2;t)\!+\!p_2(n_1,n_2;t),n_1,n_2\ne 0$, $p_s(0,0;t)=p_1(0,0;t)=p_2(0,0;t)$ (by convention) and
$p_s(t)=\sum_{n_1=0}^{\infty}\sum_{n_2=0}^{\infty}p(n_1,n_2;t)$.

The equations describing the time evolution of the probabilities $p_k(n_1,n_2;t)$ are e.g.
\begin{align}\label{eq:twos1}
&\frac{d\psone}{dt}=-(\lambda_1+\lambda_2)\psone\nonumber\\
&+\lambda_1 [p_1(n_1-1,n_2;t)+p_2(n_1-1,n_2;t-\tau_2)e^{-(\lambda_1+\lambda_2)\tau_2}]
\end{align}
for $n_1>0$ and $n_2>0$. 
The last term of this equation implements the constraint for the "not-blocked" state of the channel (see Supplementary material): 
in passing from the state \{"not-blocked", $n_1-1, n_2$, last particle entered = type 2\} to \{"not-blocked", $n_1, n_2$, last particle entered = type 1\} 
it is necessary that in the previous time  interval $(t-\tau_2,t)$ (i) no particle of type 1 or 2 enters the channel (given by the exponential term) 
and (ii) a single particle of 
type 1 enters between $t$ and $t+dt$. In analogy with Eq. (\ref{eq:cars}), this is indicative of the non-Markovian nature of the process. The evolution equation for $p_2(n_1,n_2;t)$ 
is obtained from Eq. (\ref{eq:twos1}) by symmetry.


In addition, we have to consider the time evolution of the ``boundaries" ($0,n_2$) and ($n_1,0$):
obviously $p_2(n_1,0;t)=p_1(0,n_2;t)=0$ for $n_1, n_2\ge 1$.
\begin{equation}\label{eq:twos01}
 \frac{dp_1(n_1,0;t)}{dt}=-(\lambda_1+\lambda_2)p_1(n_1,0;t)+\lambda_1p_1(n_1-1,0;t)
\end{equation}
with $n_1\ge 1$ and a corresponding equation for $p_2(0,n_2;t)$.
To complete the configuration space, one must introduce the probability that no particle is created in the time interval $(0,t)$, $p(0,0;t)$
and one has$ \frac{dp(0,0;t)}{dt}=-(\lambda_1+\lambda_2)p(0,0;t)$ with $p(0,0;0)=1$ with solution
$p(0,0;t)=e^{-(\lambda_1+\lambda_2)t}$.

As for the previous model, the solution is obtained by introducing a generating function:
\begin{equation}
 G(z_1,z_2;t)=\sum_{n_1=0}^{\infty}\sum_{n_2=0}^{\infty}z_1^{n_1}z_2^{n_2}p(n_1,n_2;t) 
\end{equation}
from which the survival probability can be found as
$p_s(t) = G(1,1;t)$.

After some calculation (see Sect.~2.1 of the SM), we obtain 
\begin{align}\label{eq:gz}
\tilde{G}&(z_1,z_2,u)=\frac{1}{1+\lambda_1+\lambda_2}\left[1+\frac{\lambda_1 z_1}{u+\lambda_2 +\lambda_1(1-z_1)}\right.\nonumber\\
&+\frac{\lambda_2 z_2}{u+\lambda_1+\lambda_2 (1-z_2)}+\frac{\lambda_1\lambda_2 z_1 z_2}{\Delta}\nonumber\\
&\left(e^{-(\lambda_1+\lambda_2+u)\tau_1}+e^{-(\lambda_1+\lambda_2+u)\tau_2}+e^{-(\lambda_1+\lambda_2+u)(\tau_1+\tau_2)}\right.\nonumber\\
&\left.\left.\left\lbrace\frac{\lambda_1z_1}{u+\lambda_2+\lambda_1(1-z_1)}+\frac{\lambda_2z_2}{u+\lambda_1+\lambda_2(1-z_2)}\right\rbrace\right)\right]
\end{align}
where
$\Delta=(u+\lambda_2+\lambda_1(1-z_1))(u+\lambda_1+\lambda_2(1-z_2))
-\lambda_1\lambda_2 z_1 z_2e^{-(\lambda_1+\lambda_2+u)(\tau_1+\tau_2)}.$
This is the principal result for the counterflow model from which most properties of interest can now be easily obtained. 
In particular, the mean survival time, $ \langle t\rangle =\tilde{p}_s(u=0)$, is
\begin{align}
 \langle t\rangle &= \frac{1}{\lambda_1+\lambda_2}\left[1+\frac{e^{-(\lambda_1+\lambda_2)\tau_1}+e^{-(\lambda_1+\lambda_2)\tau_2}}{1-e^{-(\lambda_1+\lambda_2)(\tau_1+\tau_2)}}\right.\nonumber\\
&+\frac{\lambda_1^2+\lambda_2^2}{\lambda_1\lambda_2}
\left.\frac{1}{1-e^{-(\lambda_1+\lambda_2)(\tau_1+\tau_2)}}\right]
\end{align}
The three contributions have a simple physical meaning: the first term corresponds to the situation where no species exit the passage before failure.
The second term corresponds to situations where an even number of changes of species occurs before failure, the last term to the situations where an
odd number of changes (larger than $1$) of species occurs before failure.

Note that 
 when $(\lambda_1+\lambda_2)\tau_i<<1$,
\begin{equation}\label{eq:ltcc}
 \langle t\rangle \approx \frac{1}{\lambda_1\lambda_2(\tau_1+\tau_2)}
\end{equation}
corresponding to a regime where a large number of event sequences contribute to the survival probability. 

\begin{figure}[th]
\begin{center}
\resizebox{6cm}{!}{\includegraphics{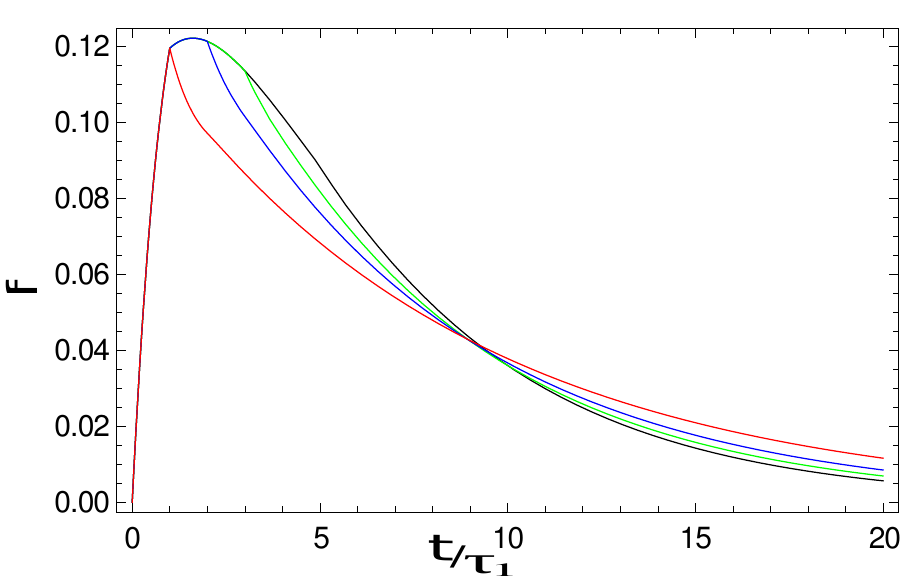}}

\end{center}
\caption{Probability of blockage, $f(t) = -dp_s(t)/dt$, as a function of time for the counterflow model. $\lambda_1=0.5, \lambda_2=0.2, \tau_2/\tau_1=1$ (red), 
$\tau_2/\tau_1=2$ (blue), $\tau_2/\tau_1=3$ (green), $\tau_2/\tau_1=5$ (black)}\label{fig:edcf}
\end{figure}

It is possible to perform a term-by-term inversion of the Laplace transform to obtain the time dependent survival probability.
For the case $\lambda_1=\lambda_2=\lambda$, $\tau_1=\tau_2=\tau$ the result is (see the SM)
\begin{align}
\label{eq:psf}
 p_s(t)&=  e^{-2\lambda t}+2\sum_{k=0}^{\infty}\theta(t-2k\tau)\left[-e^{-2\lambda t}+e^{-\lambda (t+2k\tau)}\right.\nonumber\\
&\left.\sum_{l=0}^{2k} \frac{(-1)^{l}}{l!}(\lambda (t-2k\tau))^{l}\right]+2\sum_{k=0}^{\infty}\theta(t-(2k+1)\tau)\nonumber\\
&
\left[e^{-2\lambda t}
-e^{-\lambda (t+(2k+1)\tau)}\sum_{l=0}^{2k+1}\frac{(-1)^{l}}{l!}(\lambda (t-(2k+1)\tau)^{l} \right]
\end{align}
For a given time, the solution contains a finite number of nonzero terms.
At large $t$, by using Laplace's method, we obtain
\begin{equation}
 p_s(t)\sim e^{- \left(\lambda-\frac{L_W(\lambda\tau e^{-\lambda\tau})}{\tau}\right)t}
\end{equation}
Note that when $\lambda\tau\ll 1$, $p_s(t)\sim e^{- 2\lambda^2\tau t}$
consistent with Eq. (\ref{eq:ltcc}). The average survival time is dominated by this regime.
The general solution is given in the SM and Fig. \ref{fig:edcf} 
illustrates a particular case. Note the presence of two cusps ($\tau_1\ne\tau_2$) corresponding to the two transit times.

\begin{figure}[th]
\begin{center}
\resizebox{6cm}{!}{\includegraphics{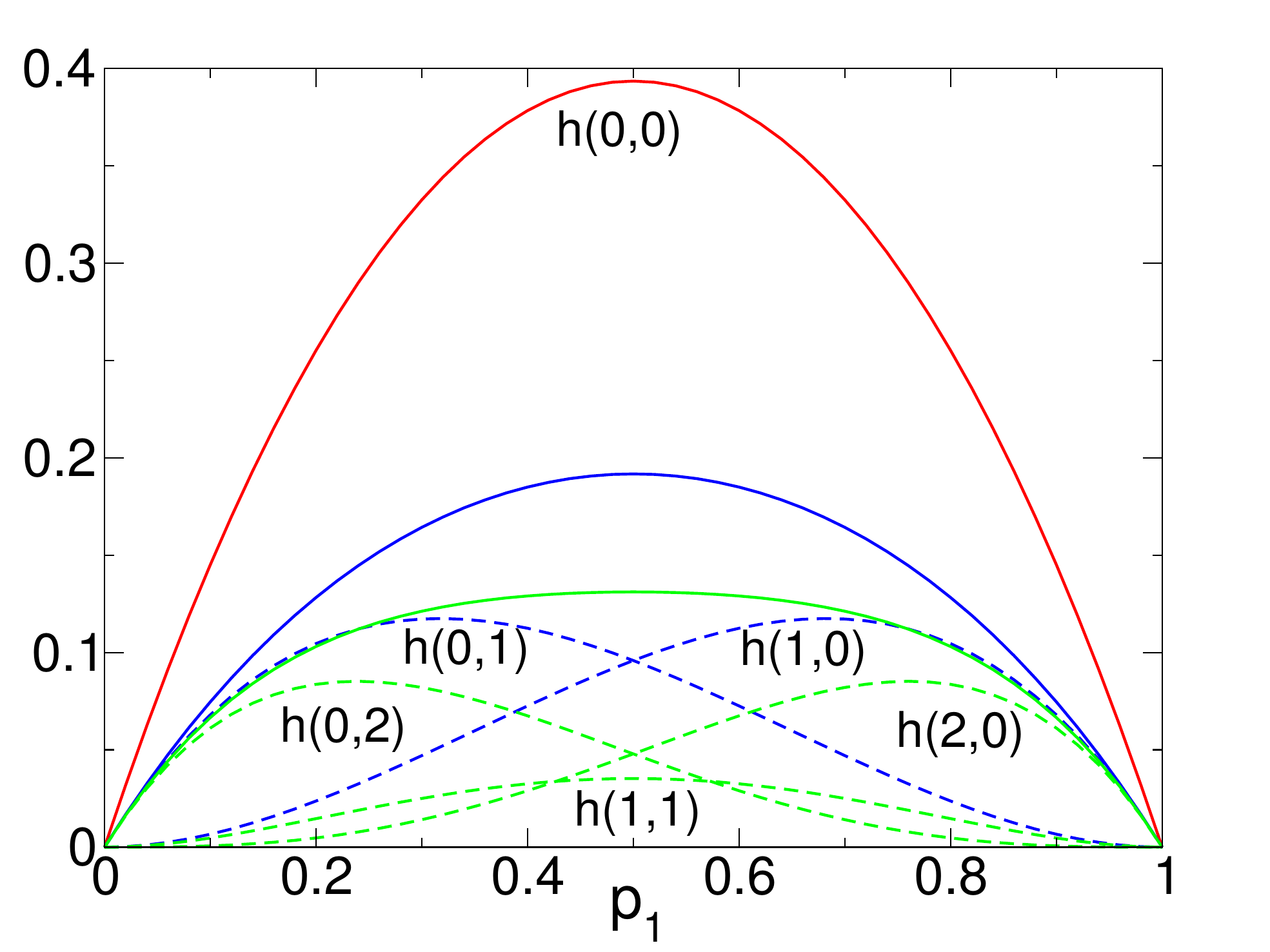}}
\end{center}
\caption{Probability of different outcomes in the counterflow model as a function of 
$p_1=\lambda_1/(\lambda_1+\lambda_2)$, $\lambda_1+\lambda_2=1$ and $\tau_1=\tau_2=1$. The solid curves indicate the probability that zero ($=h(0,0)$), 
one ($=h(1,0)+h(0,1)$) and two ($=h(2,0)+h(0,2)+h(1,1)$) 
(top to bottom) particles 
exit the channel before blockage occurs.}\label{fig:hnm}
\end{figure}

It is more difficult to obtain $h(m_1,m_2)$, the probability that $m_1$ particles of type 1 and  $m_2$ particles of type 2 exit
the passage before blockage occurs. This is because there is no simple relationship between $m_1, m_2$ and the numbers $n_1, n_2$ that 
have entered the passage as is the case in the concurrent flow model. 
However, the first few may be obtained by direct calculation: 
\begin{equation}
 h(0,0)=p_1(1-e^{-\lambda_2\tau_1})+p_2(1-e^{-\lambda_1\tau_2})
\end{equation}
\begin{equation}
 h(m_1,0)=p_1^{m_1}p_2(e^{-\lambda_2\tau_1}-e^{-(\lambda_1+\lambda_2)\tau_1}e^{-\lambda_1\tau_2}),\;\;m_1\ge 1
\end{equation}
\begin{equation}
 h(0,m_2)=p_1p_2^{n_2}(e^{-\lambda_1\tau_2}-e^{-(\lambda_1+\lambda_2)\tau_2}e^{-\lambda_2\tau_1}),\;\;m_2\ge 1
\end{equation}
and
\begin{align}\label{eq:h11}
 h(1,1)&=p_1p_2e^{-(\lambda_1+\lambda_2)(\tau_1+\tau_2)}[p_1(e^{\lambda_2\tau_2}-e^{-\lambda_2\tau_1})\nonumber\\
&+p_2(e^{\lambda_1\tau_1}-e^{-\lambda_1\tau_2})]
\end{align}
where $p_i=\lambda_i/(\lambda_1+\lambda_2),\;i=1,2$. See the SM for details.
The behavior of these functions is illustrated in Fig. \ref{fig:hnm} for the non-restrictive situation of a constant total flux $\lambda_1+\lambda_2=1$ apportioned
continuously between the left and right hand streams.  As in the concurrent flow model, the most likely result is that blockage occurs before any particles exit.

We conclude with an illustration  of the theory: 
Let us suppose that a single-track road is $0.5km$ long and 
on average $10$ cars enter each side per hour. If we 
further assume that all cars travel at a constant speed of $50km/h$
then the survival probability after $5$ minutes is $0.876$ and after $30$ minutes it is $0.436$.
The mean survival time is $36$ minutes. 

In summary we have developed stochastic models to describe the probability of blocking in diverse physical applications 
involving particulate flow.   
Both models can serve as the starting point for more refined models tailored to specific applications. 
For a filter composed of $M$ independent channels, the fraction that is active at time $t$ is just $Mp_s(t)$. 
With more effort, connected channels and reversible blocking can also be treated within the same framework.
Clustering of the particulate streams 
can be modeled using an inhomogeneous Poisson process
where the intensity is time-dependent, $\lambda(t)$.

We thank  O. B\'enichou, G. Oshanin, Ch. Pouzat, L. Signon and G. Tarjus for useful discussions.


%

\end{document}